\documentstyle[graphicx]{mn}

\title{Iron K features in the hard X-ray \xmm ~spectrum of NGC 4151}

\author[N.J.\,Schurch, R.S.\,Warwick, R.E.\,Griffiths S.\,Sembay]
{N.J.\,Schurch$^{1*}$, R.S.\,Warwick$^{2}$, R.E.\,Griffiths$^{1}$, 
S.\,Sembay$^{2}$ \\
$^{1}$Department of Physics, Carnegie Mellon University, 5000 Forbes Avenue, 
Pittsburgh, PA 15213 \\
$^{2}$Department of Physics and Astronomy, University of Leicester, 
University Road, Leicester, LE1 7RH \\
$^{*}$ schurch@andrew.cmu.edu}
\date{}

\def\exo{{\it EXOSAT\/}}
\def\gin{{\it Ginga\/}}

\def\asca{{\it ASCA\/}}

\def\xmm{{\it XMM-Newton\/}}

\def\cha{{\it Chandra\/}}
\def\bep{{\it BeppoSAX\/}}
\def\int{{\it INTEGRAL\/}}

\def\H0{{\rm ~km~s^{-1}~Mpc^{-1}}}

\def\deg{\hbox{$^\circ$}}
\def\arcm{{\hbox{$^\prime$}}}
\def\arcs{{\hbox{$^{\prime\prime}$}}}

\def\etal{et al.~\/}

\def\la{\mathrel{\hbox{\rlap{\hbox{\lower4pt\hbox{$\sim$}}}{\raise2pt\hbox{$<$}}
}}}
\def\ga{\mathrel{\hbox{\rlap{\hbox{\lower4pt\hbox{$\sim$}}}{\raise2pt\hbox{$>$}}
}}}

\def\d25{D$_{\rm 25}$}

\def\.25{0.25 keV\thinspace}

\begin{document}

\maketitle 

\begin{abstract}

Recent \xmm ~observations have measured the hard (2.5-12 keV) X-ray spectrum of the well-known Seyfert galaxy NGC 4151 with a signal-to-noise unprecedented for this source. We find that a spectral model, developed to fit previous \bep ~and \asca ~observations of NGC 4151, provides an excellent description of the \xmm ~EPIC data. The results support the view that it is the level of the continuum that is the main driver of the complex spectral variability exhibited by NGC 4151. We focus on the iron K features in the NGC 4151 spectrum. There is no requirement for a relativistically broadened iron K$\alpha$ line, in contrast to several earlier studies. The iron K$\alpha$ line profile is well modelled by a narrow Gaussian, the intensity of which varies by $\sim$25\% on timescales of about a year. There is also a strong suggestion that the cold media present in the active nucleus of NGC 4151 have an iron abundance that is at least twice the solar value.

\end{abstract}


\section{Introduction}
 
The Seyfert 1 galaxy NGC 4151 harbours one of the brightest Active Galactic Nuclei (AGN) accessible in the X-ray band and has been extensively studied by all major X-ray missions. This observational focus has revealed the X-ray spectrum of NGC 4151 to be a complex mixture of emission and absorption components, originating at a variety of locations from the innermost parts of the putative accretion disk, out to the extended narrow-line region of the galaxy. The X-ray spectrum emanating from the active nucleus is dominated by an intrinsic X-ray to $\gamma$-ray continuum probably produced by the thermal Comptonization of soft seed photons ({\it e.g.} Zdziarski \etal 1996, 2000). There is also evidence in the X-ray spectrum for continuum reprocessing features, in the form of a Compton reflection continuum and emission/absorption features imprinted by iron (Yaqoob \etal 1993; Zdziarski \etal 2002; Piro \etal 2002). Below $\sim$ 5 keV the hard continuum is strongly cut-off by photoelectric absorption in a substantial ($N_{H}$$\sim$10$^{23}$ cm$^{-2}$) line-of-sight gas column density (Holt \etal 1990; Yaqoob \etal 1993; Weaver \etal 1994b; Schurch \& Warwick 2002). Additional soft X-ray emission becomes apparent below $\sim$2 keV and dominates the spectrum below $\sim$1 keV (Weaver \etal 1994a,b, Warwick, Smith \& Done 1995). This soft emission has been spatially resolved by \cha (Ogle \etal 2000; Yang, Wilson \& Ferruit 2001) and is co-spatial with the O III ionization cones identified in {\it Hubble Space Telescope} observations (Ogle \etal 2000) implying a strong association between the soft X-ray emission and the extended narrow-line region (ENLR) of is galaxy.

Some earlier studies, based on \asca ~observations, concluded that the profile of the iron K${\alpha}$ emission line is complex and may be composed of an intrinsically narrow component plus a relativistically broadened line feature (Yaqoob \etal 1995; Wang \etal 2001). Furthermore, the broad line profile has been reported to be variable on timescales of 10$^{4}$s, corresponding to an emitting region of $<$0.02 AU, suggesting an origin close to the supermassive black hole presumably in the inner regions of a putative accretion disk (Wang \etal 2001). However, there are alternative interpretations of the broadband \asca ~spectra that do not require the presence of an extremely broad iron line in NGC 4151 (Schurch \& Warwick 2002; Takahashi \etal 2002).

In this paper we focus on the iron K features present in the extremely high signal-to-noise spectra of NGC 4151 recorded by the EPIC CCD cameras on \xmm. Our analysis makes use of the spectral ``template'' model, developed by Schurch \& Warwick (2002) to interpret earlier \asca ~and \bep ~observations.

\section{The \xmm ~observations}

NGC 4151 was observed by \xmm ~three times during the period December, 21$^{st}$-23$^{rd}$ 2000 (orbit 190). In all three observations the EPIC MOS and PN cameras (Turner \etal 2001; Str{\"u}der \etal 2001) were operated with the medium filter in place. The first observation ($\sim$33 ks) was carried out with both CCD systems in {\it Small Window Mode} ({\it SWM}), whereas the remaining two observations ($\sim$23 ks and $\sim$63 ks respectively) employed {\it Full Window Mode} ({\it FWM}). The events recorded in {\it FWM} were screened with the \xmm ~Science Analysis Software (SAS v5.2) to remove known hot pixels and other data flagged as bad. The {\it SWM} observation was processed with a local developers version of the SAS, which included a patch with the most up-to-date correction for the PN charge transfer inefficiency (this correction is now included in the latest public release version of the SAS, v5.3). The data were processed using the latest CCD gain values and only X-ray events corresponding to patterns 0-12 in the MOS cameras and 0-4 in the PN camera were accepted. An examination of the temporal variation in count rate from the full available field (but excluding the bulk of the NGC 4151 contribution) in all three observations revealed a single background flaring event during the first observation, which was screened from subsequent analysis. The resulting total effective exposure times were 110 ks and 91 ks for the MOS and PN instruments respectively.
 
An investigation into the impact of pile-up on the observation showed that the effect was negligible in almost all cases, largely due to the low flux state of NGC 4151 during the observation period. The exception to this is the {\it FWM} MOS observation, which shows marginal pile-up above 8 keV. The best-fit spectral parameters and $\chi^{2}$ values quoted in this paper are for spectral fits in which the MOS {\it FWM} data above 8 keV have been excluded.

\section{The X-ray light-curve}

A source light-curve was extracted for the MOS cameras from each individual, screened observation. Source counts were taken from within a 100\arcs x100\arcs ~box for the {\it SWM} observation and a 2\arcm ~radius circle for the {\it FWM} observations. Background subtraction was not applied since the background contributes no more than a few percent of the total signal.

\begin{figure}
\centering
\begin{minipage}{85 mm} 
\centering 
\hbox{\centering\includegraphics[height=8.4 cm, angle=270]{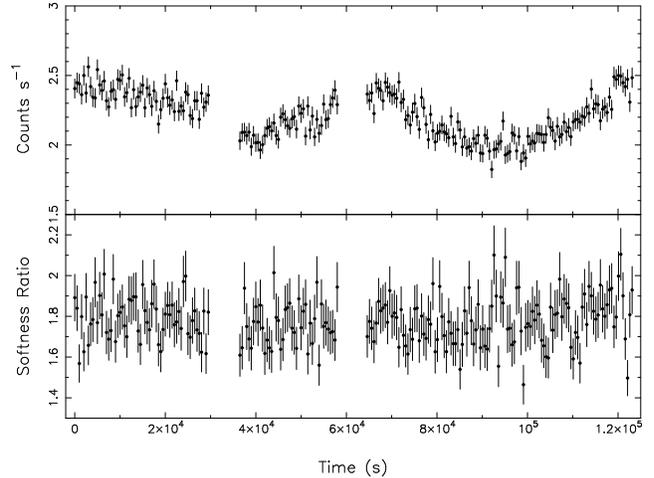}}
\caption{{\it Upper Panel}: The 2-10 keV light-curve from the co-added \xmm ~EPIC MOS detectors. The light-curve is binned to a resolution of 500 seconds and covers the three individual observations (separated by the data gaps). {\it Lower Panel}: The variation of the 2-5 keV/5-10 keV softness ratio during the observations.}
\label{lc}
\end{minipage}
\end{figure}

The 2-10 keV light-curve (Fig. \ref{lc}, {\it Upper Panel}) shows slow drifts in the X-ray flux of NGC 4151 on timescales longer than $\sim$10$^{4}$ s, behaviour which is very characteristic of this source ({\it e.g.} Yaqoob \& Warwick 1991; Yaqoob \etal 1993). The largest flux variation seen was an increase of $\sim$25\% over the last $\sim$2$\times$10$^{4}$ s of the observation. Previous observations of NGC 4151 have revealed complex spectral changes associated with long-timescale flux variations (Schurch \& Warwick 2002). However, the 2-5 keV/5-10 keV softness ratio measured during the \xmm ~observations (Fig. \ref{lc}, {\it Lower Panel}) remains fairly constant, with just a hint of spectral softening towards the end of the observation correlated with the flux increase noted earlier.

\section{Spectral analysis}

In the following analysis we consider the background-subtracted spectrum of NGC 4151 averaged over the three \xmm ~observations. For the MOS spectra, source counts were extracted from the same source regions as employed for the light-curves. In the {\it FWM} observations, the background was taken from a region offset from the source response but on the same central chip (MOS 1 and 2 cameras). The {\it FWM} background was also used to correct the {\it SWM} observation. For all the MOS observations the total background amounted to less than 3\% of the source plus background count rate.

For the PN camera we used a 2\arcm ~radius source cell for the {\it FWM} observations with an offset background region located on the same chip as the source image. For the {\it SWM} observation we derived the source spectrum using a 4\arcm x2\arcm box and the background from a similar size region offset from the source but on the same chip. In this latter case the contamination of the background region by the wings of the source response resulted in a background count rate amounting to $\sim$8\% of the source plus background signal. The loss of source flux in the various cases is accounted for both in the spectral fitting and in the quoted fluxes.

For convenience the spectra from the two {\it FWM} observations were co-added to produce a single source spectrum and a single background spectrum for each instrument. The resulting spectra were binned to a minimum of 20 counts per spectral channel, in order to apply $\chi^{2}$ minimisation techniques in the spectral fitting process. Here we restrict the following analysis to the 2.5-12 keV bandpass. A study of the complex soft X-ray spectrum of NGC 4151 based on \xmm ~EPIC and RGS data is in progress and will be presented elsewhere.

We adopt the spectral ``template'' model described by Schurch \& Warwick (2002), which includes the following emission components:

\begin{enumerate}

\item An absorbed power-law continuum with a fixed photon index, $\Gamma$=1.65, and a high energy break at 100 keV. The normalization of the power-law, A$_{1}$, is a free parameter in the modelling.

\item A neutral Compton reflection component (represented by the XSPEC model {\bf pexrav}; Magdziarz \& Zdziarski 1995) with only the reflection scaling factor, $R$, as a free parameter. Specifically, the normalisation and spectral index of the illuminating component were fixed at the same values as those for the underlying power-law continuum. In addition cos $i$ was set at 0.5 and the metal abundance in the reflector was taken to be the solar value.

\item An iron K$\alpha$ emission line with a Gaussian profile, with the line intensity I$_{K\alpha}$, the line energy E$_{K\alpha}$ and the intrinsic line width $\sigma_{K\alpha}$ all free parameters.

\end{enumerate}

The absorption of the power-law continuum (which is known to be complex in NGC 4151) is represented as the product of two absorption components, namely a partially photoionized ({\it i.e.} warm gas) component with column density N$_{H,warm}$ and a cold gas column N$_{H,cold}$. The state of the warm absorber is governed by the ionization parameter $\xi$ defined as $\xi$=L$_{ion}$/nr$^{2}$, where L$_{ion}$ is the source luminosity in the 0.0136-13.6 keV bandpass (erg s$^{-1}$), $n$ is the number of hydrogen atoms/ions in the gas (cm$^{-3}$) and $r$ is the distance from the central source to the inner edge of the warm cloud (cm). The value of $\xi$ is a free parameter in the model, but the column densities of the two absorption components are fixed at the values used by Schurch \& Warwick (2002), namely N$_{H,warm}$=2.4$\times$10$^{23}$ cm$^{-2}$ and N$_{H,cold}$=3.4$\times$10$^{22}$ cm$^{-2}$. For further details of the photoionization modelling see Schurch \& Warwick (2002) and Griffiths \etal (1998). The adopted spectral model also includes absorption arising in the line-of-sight column density through our own Galaxy (N$_{H,Gal}$=2$\times$10$^{20}$ cm$^{-2}$), applied to all the emission components. 

\begin{figure}
\centering
\begin{minipage}{85 mm} 
\centering \includegraphics[height=8.4 cm, angle=270]{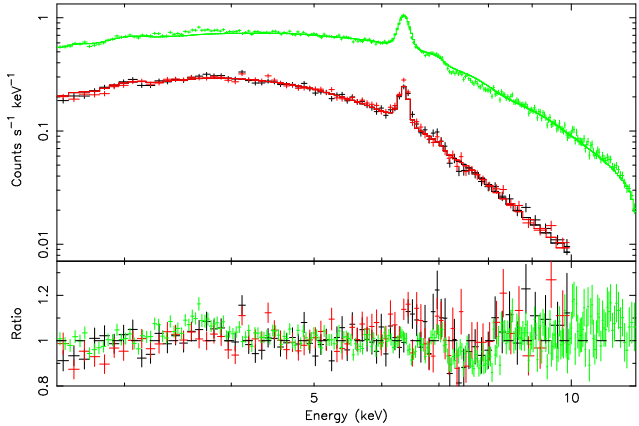}
\caption{{\it Upper Panel}: The 2.5-12 keV X-ray spectra measured by the EPIC instrument on \xmm. The MOS data, shown here in black (MOS 1) and red 
(MOS 2), are from the {\it SWM} observation. The PN data, shown here in green, are from the {\it FWM} observation. The solid line shows the best-fitting spectral ``template'' model (Table~\ref{modelfits}, Model 1). {\it Lower Panel}: The ratio of the data to the model prediction.}
\label{4151spect_1}
\end{minipage}
\end{figure}

In summary, the above spectral model requires a total of just six free parameters relating to the continuum and reflection normalisations, the iron K$\alpha$ line properties and the ionization state of the warm column. However, in the simultaneous fitting of the {\it SWM} and {\it FWM} spectra for each detector (MOS 1, MOS 2, PN), we also employ five constant factors allowing for up to a $5\%$ relative flux scaling between the different spectral datasets. Although the resulting best-fit was formally unacceptable, with $\chi^{2}$=5193 for 4756 degrees of freedom (d.o.f), the remaining residuals were largely at the few percent level, close to the limit of the calibration uncertainty of the \xmm EPIC cameras. Including a 1.5\% systematic error to allow for these calibration uncertainties (S. Sembay, private communication) reduces the $\chi^{2}$ to 5049. By way of illustration, Fig.~\ref{4151spect_1} shows three of the six available EPIC spectra datasets along with the overall best-fitting model and the data/model fit residuals. The corresponding best-fitting parameter values are listed in Table \ref{modelfits} as Model 1. The quoted errors (here as elsewhere in this paper) are for a 90\% confidence level as defined by a $\Delta$$\chi$$^{2}$=2.71 criterion ({\it i.e.}, assuming one interesting parameter).

\begin{table}
\centering
\begin{minipage}{85 mm} 
\centering
\caption{The best-fitting spectral parameters}
\begin{tabular}{lcc}
Parameter & Model 1 & Model 2 \\ \hline
R & 1.95$^{+0.11}_{-0.25}$ & 1.93$^{+0.14}_{-0.23}$ \\
log $\xi$ & 2.634$^{+0.003}_{-0.008}$ & 2.624$^{+0.004}_{-0.006}$ \\ 
A$_{pl}$$^{a}$ & 1.28$^{+0.03}_{-0.01}$ & 1.31$^{+0.03}_{-0.02}$ \\
E$_{K\alpha}$$^{b}$ & $6.393^{+0.003}_{-0.002}$ & 6.393$^{+0.003}_{-0.002}$ \\ 
$\sigma_{K\alpha}$$^{b}$ & 0.039$^{+0.005}_{-0.006}$ & 0.033$^{+0.006}_{-0.005}$ \\
I$_{K\alpha}$$^{c}$ & $1.32^{+0.03}_{-0.03}$ & 1.26$^{+0.04}_{-0.04}$ \\
E$_{K,Edge}$$^{b}$ & - & 7.11$^{+0.01}_{-0.01}$\\
$\tau_{K, Edge}$ & - & 0.12$^{+0.02}_{-0.02}$\\
 & & \\
$\chi^{2}$ & 5049 & 4859 \\
d.o.f & 4756 & 4754 \\ \hline
\multicolumn{2}{l}{\scriptsize $^{a}$ 10$^{-2}$ photon keV$^{-1}$ cm$^{-2}$ s$^{-1}$} & \\
\multicolumn{1}{l}{\scriptsize $^{b}$ keV} \\
\multicolumn{2}{l}{\scriptsize $^{c}$ 10$^{-4}$ photon cm$^{-2}$ s$^{-1}$} & \\
\end{tabular}
\hspace{-1 cm}
\label{modelfits}
\end{minipage}
\end{table}

Remarkably, the spectral template model provides an excellent description of the \xmm ~spectra with only minor adjustments to a limited number of free parameters. Given the very high quality of the EPIC spectra (the PN and the combined MOS 1/2 spectra contain in excess of 340,000 and 240,000 counts in the hard X-ray band, respectively!), it would be surprising if further subtle spectral features were not present in the data. In fact, inspection of the data/model residuals (see Figure \ref{4151spect_1}, {\it Lower Panel}) do suggest a systematic deficit of counts between 7 and 8 keV. This is interpreted as a need for extra absorption over and above the current model prediction, which in terms of the spectral modelling can be accommodated by the addition of an extra absorption edge, with the edge energy and optical depth as free parameters. Also at this stage a contribution from (neutral) iron K$\beta$ emission is included; for this purpose a Gaussian line was added at a rest frame energy of 7.058 keV, with the same intrinsic width as the iron K$\alpha$ line and at 10\% of the K$\alpha$ intensity ({\it i.e.} the appropriate branching ratio).

\begin{figure}
\centering
\begin{minipage}{85 mm} 
\centering 
\includegraphics[width=5.5 cm, angle=270]{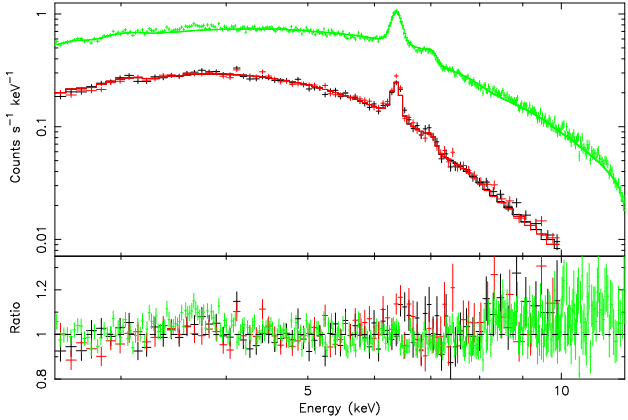}
\caption{{\it Upper Panel}: The 2.5-12 keV X-ray spectra measured by the EPIC instrument on \xmm. The MOS data, shown here in black (MOS 1) and red (MOS 2), are from the {\it SWM} observation. The PN data, shown here in green, are from the {\it FWM} observation. The solid line shows the best-fitting spectral model (Table~\ref{modelfits}; Model 2). {\it Lower Panel}: The ratio of the data to the model prediction.}
\label{4151spect_2}
\end{minipage}
\end{figure}

The result of these modifications was a significantly improved fit ($\chi^{2}$=4859 for 4754 d.o.f). The model fit to the data is shown in Fig.~\ref{4151spect_2} and the best-fitting parameter values for this revised model are listed as Model 2 in Table~\ref{modelfits}. The null-hypothesis probability of this latter model (0.14) indicates that this is a reasonable representation of the spectral data to within the limits of the instrument calibration.

\section{Discussion}
 
On the basis of the models in Table \ref{modelfits}, the absorption corrected flux of NGC 4151 measured by \xmm ~was 5.8$\times$10$^{-11}$ erg s$^{-1}$ (2-10 keV). This represents a very faint state of the source as judged against previous extensive X-ray monitoring by missions such as \exo, \gin, \bep ~and \asca ~(Yaqoob \etal 1993; Piro \etal 2002; Schurch \& Warwick 2002).

A number of points follow from a detailed comparison of the parameter values in Table \ref{modelfits} (Model 2) with those quoted in Schurch \& Warwick (2002). For example, the ionization parameter pertaining to the warm absorbing medium remains high (log $\xi$=2.62) despite the decline in the level of the hard power-law continuum by, for example, almost a factor 4 compared with the January 1999 \bep ~observation. This may be further evidence for long recombination timescales in the warm absorber as discussed in detail by Schurch \& Warwick (2002). In actual fact, the value of the ionization parameter derived from the \xmm ~observation falls at the high end of the range recorded in the \asca ~long-look observation of May, 2000 ($\xi$=2.48-2.65), although in the latter case the underlying continuum was a factor of 1.5-4 brighter. The rather high value of the reflection parameter (R$\approx$2) derived from the \xmm ~spectra may be explained if the continuum source has fairly recently and/or temporarily entered a low-flux period, whereas the Compton reflection has remained high due to the time lag inherent in a component originating in material relatively distant from the central source of ionizing radiation. For example, the bulk of the Compton reflection may be produced from the far side of a putative molecular torus in NGC 4151, in which case the response time would be measured in years. The {\it measured flux} of the Compton reflection signal is a factor of 1.5$^{+1.2}_{-0.6}$ times higher than observed during the January 1999 \bep ~observation; this encompasses the possibility of little or no change in the Compton reflection signal over a two year interval.

\subsection{The profile of the iron K$\alpha$ line}

Above we model the iron K$\alpha$ emission observed in the EPIC ~spectra as a single, ``narrow'' (see \S6.3) Gaussian component. The line energy is consistent with the fluorescence of neutral or near-neutral iron (taking into account up to 5 eV systematic error in the energy calibration). The line energy measured previously by \asca ~was not sufficiently well constrained to rule out the possibility that the iron K$\alpha$ line is seen in purely scattered light (the
first-order peak due to Compton scattering is lower by $\sim$80 eV than the zeroth-order peak at 6.4 keV; Jourdain \& Roques 1995). The iron K$\alpha$ line energy measured in the \xmm ~spectra rules out this possibility; the dominant line component is seen directly rather than as a result of a strong scattering of the iron Ka line in the extended medium above and below the accretion disk. Including a second Gaussian line component with a fixed energy of 6.4 keV and a significantly broader profile ({\it e.g.} $\sigma \approx 0.2$ keV; Yaqoob \etal 2001) does not improve the overall fit; the upper limit on the flux in such a component is 1.0$\times$10$^{-5}$ photons cm$^{-2}$ s$^{-1}$. Fixing the energy of the broad Gaussian line at 6.2 keV (as found in previous studies - Yaqoob \etal 1995) results in a very weak line (the best-fit normalisation is an upper limit of 5.1$\times$10$^{-6}$ photons cm$^{-2}$ s$^{-1}$) and does not improve the fit significantly ($\Delta\chi^{2}$=2 for 2 d.o.f). Substituting the line emission expected from an accretion disk (the disk-line model in XSPEC, Fabian \etal 1989), for the broad Gaussian component in this model, also results in a very weak line (best-fit normalisation is an upper limit of 2.75$\times$10$^{-6}$ photons cm$^{-2}$ s$^{-1}$) that does not provide any improvement in the quality of the fit ($\Delta\chi^{2}$=0.2 for 2 d.o.f). Replacing the two line profiles in these fits with a single line associated with emission from the accretion disk results in a similar $\chi^{2}$ to that obtained with a narrow line. However, in this case the derived inner radius of the accretion disk is r$_{i}$, $\sim$1000 r$_{g}$ (q$\approx$2, $\theta\approx$45\deg), which results in a disk-line profile that closely resembles a narrow Gaussian line profile.

 In summary, the high significance \xmm ~spectra reveal only a narrow iron K$\alpha$ line with no evidence for the presence of a strong relativistically broadened iron K$\alpha$ feature. This is contrary to the earlier conclusions of Yaqoob \etal (1995) and Wang \etal (1999, 2001), who on the basis of \asca ~spectra identified broad iron K$\alpha$ features in the NGC 4151 spectrum. However, we note that the presence of a broad line was not required in the spectral modelling of the final \asca ~long-look observation of NGC 4151 (Schurch \& Warwick 2002; Takahashi \etal 2002). It remains possible that a relativistically broadened line appears in the X-ray spectrum of NGC 4151 only intermittently or for particular `states' of the source. Further monitoring of NGC 4151 with \xmm ~and \cha ~should help clarify whether this is the case.

\subsection{The iron K$\alpha$ line flux}
\label{ikaflux}

The measured flux in the narrow iron $K\alpha$ line is $\sim$1.3$\times$10$^{-4}$ photons cm$^{-2}$ s$^{-1}$, significantly lower than that measured in earlier \asca, \bep ~and \gin ~observations, where the line flux was more typically $\sim$2.2$\times$10$^{-4}$ photons cm$^{-2}$ s$^{-1}$; (Yaqoob \& Warwick 1991; Yaqoob \etal 1993; Schurch \& Warwick 2002).
The present measurement is also somewhat lower than that reported in recent \cha ~observations (1.8$\pm$0.2$\times$10$^{-4}$ photons cm$^{-2}$ s$^{-1}$; Ogle \etal 2000). Ogle \etal (2000) state that \cha ~spatially resolves 65$\pm$9\% of the iron line emission, placing its origin in the ENLR at a distances of up to $\sim$500 pc from the continuum source. Thus an iron K$\alpha$ line flux of up to 1.2$\times10^{-4}$ photons cm$^{-2}$ s$^{-1}$ could be attributable to the ENLR, a value comparable to that measured in the current \xmm ~observations. The implication is that a component of the iron K$\alpha$ line produced close to the central engine in NGC 4151 almost completely disappeared over the $\sim$10 month time interval between the \cha ~and \xmm ~observations. However, we caution that this intriguing scenario may not be valid. The line intensity given by Ogle \etal (2000) combined with the exposure time and HETG effective area indicate $\sim$160 photons in the line. The central $\sim$1\arcs ~of the cross-dispersion profile integrated over the energy interval containing the line contains the majority of the photons from both line and continuum (whether extended or not). To determine whether any of the line emission is extended to a given level of confidence, one needs to show that the cross-dispersion profile outside of the central $\sim$1\arcs ~lies above the background noise. Ogle \etal (2000) do not give a confidence level to their claim that the 1-3\arcs ~interval either side of the peak of the cross-dispersion profile contains extended iron line emission, however a brief calculation of the cross-dispersion profile for the number of photons detected in the \cha ~grating observation indicates that strongly significant ({\it i.e. $>$4$\sigma$ significance) spatially extended iron K$\alpha$ line emission {\it cannot} be detected in this relatively short \cha ~observation. If all the iron K$\alpha$ line flux instead originates close to the central engine, then the line flux from this vicinity would only need to decrease by $\sim$35\% between the \xmm ~and \cha ~observations. The intensity of the neutral iron K$\alpha$ line in the \xmm ~spectra is consistent with all the line flux originating in a nearly Compton thick molecular torus (the EQW with respect to the Compton reflection continuum, is ~1.06 keV). The decrease in line flux observed between the \bep ~and \xmm ~observations is, however, in marked contrast with the behavior of the Compton reflection component, which remains at a similar flux-level or, if anything, is slightly stronger during the \xmm ~observation than it was in the January 1999 \bep ~ observation. This decoupling of the iron K$\alpha$ and the reflection continuum suggests that the additional iron K$\alpha$ line flux observed in the earlier observations is formed in optically-thin, cool clouds located within about a light year of the central source (i.e. closer than the material responsible for the Compton reflection continuum). In this setting the decline of the iron Ka line flux presumably coincides with the central source entering a fairly prolonged low-flux period, sampled only latterly by the \xmm ~observations.

The variability constraints suggest that a significantly component of the iron K$\alpha$ line may be associated with the complex absorbing medium, including the warm absorber. Schurch \& Warwick (2002) note that the predicted range of iron ionization states in the warm medium, determined from a long-look \asca ~observation, is Fe XVI-Fe XXI (for which the associated K$\alpha$ line energy is 6.4-6.6 keV). Since the line energy measured by \asca ~(E$_{K\alpha}$=6.395$\pm$0.012 keV), excludes much of this range, it was concluded that the bulk of the line flux derives from less strongly photoionized material located further from the nucleus than the warm medium. Comparison of the \asca ~results with the present \xmm ~observations suggests the narrow iron K$\alpha$ line may typically be composed of a relatively constant component amounting to $\sim$1.2$\times$10$^{-4}$ photons cm$^{-2}$ s$^{-1}$ plus a more variable component of comparable magnitude. In such a situation the \xmm ~line measurements clearly relate predominantly to the first of these two origins. The line equivalent width ($\sim$175 eV referenced to the observed power-law but $\sim$60 eV for a more typical continuum level) and energy derived from the \xmm ~observations are consistent with an origin for the ``constant'' line component in reflection from a neutral of lightly photoionized medium located outside the immediate proximity of the nucleus {\it i.e.} r$>$1 pc. On the other hand the variable line emission component may still be associated with the warm absorbing medium provided the dominant ionization state is closer to Fe XVI than Fe XXI (which could apply if our line of sight samples the high end of the ionization distribution in the warm medium). The typical equivalent width of $\sim$50 eV for the variable component coupled with the estimate, N$_{H,warm}$$\approx$2$\times$10$^{23}$ cm$^{-2}$, then implies that the warm medium subtends an angle of $\sim$$\pi$/A$_{Fe}$ steradians as viewed from the central source, where A$_{Fe}$ is the iron abundance relative to solar.

\subsection{The intrinsic width of the narrow iron K$\alpha$ line}

The global best-fit to the \xmm ~data is obtained for an intrinsic line width of the ``narrow'' iron K$\alpha$ line, $\sigma_{K\alpha}$=33$^{+6}_{-5}$ eV. However, continuing uncertainties in the CTI corrections for the current EPIC response matrices will impact on this measurement. An analysis restricted to the EPIC MOS spectra provides a good check on the value of the line width, despite the marginally worse spectral resolution at 6.4 keV, since the calibration is relatively better known for the MOS cameras compared with the PN camera. This analysis used epoch-specific response matrices, incorporating time dependent CTI corrections (S. Sembay, private communication) and yielded an intrinsic line width of $\sigma_{K\alpha}$=51$^{+7}_{-7}$ eV (Note: instrument response matrices with a non-standard 1 eV energy scale were used for this analysis). By way of comparison the effective MOS instrument resolution at the observation epoch was $\sigma$$\sim$70 eV (FWHM$\sim$165 eV). At face value the EPIC spectra suggest an intrinsic width of between 3000-6000 km s$^{-1}$ (FWHM) for the iron K$\alpha$ line. In contrast, Ogle \etal (2000) report, on the basis of \cha ~HETG observations, that ``the narrow core of this line is unresolved, with FWHM=1800$\pm$200 km s$^{-1}$''. Clearly this cautions against over-interpreting the EPIC line widths, particularly since the measurements are close to the limit of the present calibration.

What might be the origin of such line broadening if confirmed? One consideration is that the iron K$\alpha$ line is in fact a doublet with components at 6.404 (K$\alpha_{1}$) and 6.391 (K$\alpha_{2}$) keV. However, we find that modelling the iron K$\alpha$ emission feature with two Gaussian components with a fixed 13 eV energy separation and flux ratio of 2:1 (as per the branching ratio) has little impact on the above results. A further possibility is that the K$\alpha$ line appears broadened because it represents a blend of ionization states. However, the measured line energy is fully consistent with that of neutral iron at 6.400 keV (representing the centroid of the K$\alpha_{1}$ and K$\alpha_{2}$ components) and we find that adding non-neutral line components fails to replicate the low energy wing of the line. Finally, the line broadening could be due to Doppler shifts, which would place the origin of the K$\alpha$ line observed by \xmm ~squarely in the broad line region of the galaxy. Unfortunately this description is then at odds with our earlier conclusion that the relatively ``constant'' component of the line emission (observed by \xmm) probably originates in reflection from lightly photoionized matter beyond $\sim$1 pc from the nucleus.

\subsection{The additional iron edge - a clue to the iron abundance.}

The best-fit spectral model includes iron edge features imprinted on the direct continuum by the neutral and warm absorbers and also arising in the Compton reflection component. In each case solar metal abundances are assumed. However, as detailed earlier the high signal-to-noise EPIC spectra also require the inclusion of an additional absorption above 7 keV suggestive of an overabundance of iron in the absorbing and/or reflecting media. 
Modelling the hard X-ray spectrum in the region of the iron edge with a simple model incorporating a power-law continuum and a single absorption edge yields an edge energy of 7.11$^{+0.01}_{-0.01}$ keV and a total optical depth of $\tau_{K,Edge}$=0.23$^{+0.02}_{-0.02}$. The depth of this edge overpredicts the observed iron line flux by a factor of $\sim$2.5, implying a covering fraction for the absorbing material of $\sim$40\%. The derived edge energy is consistent with absorption by neutral iron (with the upper limit of 7.13 keV, including a systematic uncertainty of 5 eV, formally excluding even Fe II). This rules out an association of the excess absorption with the warm absorber, for which the predicted range of edge energy is 7.8-8.3 keV (Fe XVI-Fe XXI). If we interpret the observed optical depth in the edge ($\tau_{K,Edge}$$\approx$0.12) as due to an extra-solar abundance of iron in the cold absorber we obtain, A$_{Fe}$$\sim$3.6, whereas its association with the Compton reflector would imply A$_{Fe}$$\sim$2.3. Alternatively if we allow an overabundance in {\it both} the cold absorber and the reflector then A$_{Fe}$$\sim$2 is required. Previous studies have also reported evidence for an overabundance of iron in NGC 4151, for example Yaqoob \etal 1993 quote a canonical value of A$_{Fe}$$\sim$2.5 times solar, based on the average of many \gin ~measurements.

\section{The Conclusions}

Recent observations with the EPIC cameras on \xmm ~have provided a very high signal-to-noise measurement of the hard (2.5-12 keV) X-ray spectrum of the archetypal Seyfert galaxy NGC 4151. Remarkably, we find that the spectral ``template'' model developed by Schurch \& Warwick (2002) to fit earlier \bep ~and \asca ~observations, also provides an excellent description of the \xmm ~data, with only minor modifications. 

This supports the view that the complex spectral variability exhibited by NGC 4151 over timescales longer than a few days is the product of changes in the source, driven predominantly by the level of the underlying continuum. The ingredients of the spectral model are surprisingly simple. An underlying power-law continuum is absorbed below $\sim$5 keV by a combination of warm and cold gas along the line-of-sight but supplemented at higher energy by Compton-reflection. The parameters specifying the continuum slope, the column densities and the Compton reflection flux remain essentially constant. There is no requirement for a relativistically broadened iron K$\alpha$ line feature, but the intensity of the narrow component definitely varies on long (years) timescales. There is also a strong suggestion that either the cold absorbing and/or reflecting media have an iron abundance that is twice solar.

Further development of this picture of the hard X-ray properties of NGC 4151 will be possible through future approved \xmm ~observations aimed at monitoring spectral changes in the 2.5-12 keV band. With the addition of a simultaneous \int ~observation it will be possible to test the assumption that it is the level of the broadband hard X-ray continuum which governs such changes.

\section{Acknowledgments}

We thank the referee, Dr Tahir Yaqoob, for his helpful comments on an earlier draft of this paper. NJS gratefully acknowledges the financial support from PPARC. REG acknowledges the support of NASA grant NAG5- 9902. It is a pleasure to thank Keith Arnaud for his help with XSPEC. This work is based on observations obtained with XMM-Newton, an ESA science mission with instruments and contributions directly funded by ESA Member States and the USA (NASA). The authors wish to thank the {\em XMM-Newton team}, and particularly the EPIC calibration teams, for the hard work and dedication that underlies the success of the mission. This research has made extensive use of NASA's Astrophysics Data System Abstract Service.


\begin{thebibliography}{}
\bibitem[\protect \astroncite{Fabian \etal}{1989}]{Nschurch-C2:fab89} 
Fabian A.~C., Rees M.~J., Stella L., White N.~E., 1989, MNRAS, 238, 729.
\bibitem[\protect \astroncite{Griffiths \etal}{1998}]{Nschurch-C2:gri98}
Griffiths R.~G., Warwick R.~S., Georgantopoulos I., Done C., Smith D.~A, 1998, MNRAS, 298, 1159
\bibitem[\protect \astroncite{Haardt \& Maraschi}{1991}]{Nschurch-C2:haa91}
Haardt F., Maraschi L., 1991, 380, L51
\bibitem[\protect \astroncite{Holt \etal}{1980}]{Nschurch-C2:hol80}
Holt S.~S., Mushotzky R.~F., Boldt E.~A., Serlemitsos P.~J., Becker R.~H., Szymkowiak A.~E., White N.~E., 1980, ApJL, 241, L13
\bibitem[\protect \astroncite{Magdziarz \& Zdziarski}{1995}]{Nschurch-C2:mag95} 
Jourdain E., Roques J.~P., 1995, ApJ, 440, 128.
\bibitem[\protect \astroncite{Jourdain \& Roques}{1995}]{Nschurch-C2:jr95}
Magdziarz, P., Zdziarski, A.~A., 1995, MNRAS, 273, 837.
\bibitem[\protect \astroncite{Morrison \& McCammon}{1983}]{Nschurch-C2:mor83}
Morrison R., McCammon D., 1983, ApJ, 270, 119
\bibitem[\protect \astroncite{Nicastro \etal}{1999}]{Nschurch-C2:Nic99}
Nicastro F., Fiore F., Perola G. C., Elvis M., 1999, ApJ, 512, 184
\bibitem[\protect \astroncite{Ogle \etal}{2000}]{Nschurch-C2:Ogl00}
Ogle P.~M., Marshall H.~L., Lee J.~C., Canizares C.~R., 2000, ApJL, 545, L81
\bibitem[\protect \astroncite{Petrucci \etal}{2000}]{Nschurch-C2:pet00}
Petrucci P.~O., Haardt F., Maraschi L., Grandi P., Matt G., Nicastro F., Piro L., Perola G.~C., De Rosa A., 2000, ApJ, 540, 131
\bibitem[\protect \astroncite{Ross \& Fabian}{1993}]{Nschurch-C2:Ros93}
Ross R. R., Fabian A. C., 1993, MNRAS, 261, 74.
\bibitem[\protect \astroncite{Ross \& Fabian}{1993}]{Nschurch-C2:Ros99}
Ross R. R., Fabian A. C., Young A. J., 1999, MNRAS, 306, 461.
\bibitem[\protect \astroncite{Schurch \& Warwick}{2002}]{Nschurch-C2:Sch02}
Schurch N.~J., Warwick R., 2002, MNRAS, in press.
\bibitem[\protect \astroncite{Str{\"u}der \etal}{2001}]{Nschurch-C2:str01} 
Str{\"u}der L., Briel, U., Dennerrl, K. et al. 2001, A\&A, 365.
\bibitem[\protect \astroncite{Turner \etal}{2001}]{Nschurch-C2:Tur01} 
Turner M.J.L., Abbey, A.F., Arnaud, M. et al. 2001, A\&A, 365.
\bibitem[\protect \astroncite{Takahashi \etal}{2002}]{Nschurch-C2:Tak02} 
Takahashi K., Inoue H., Dotani T., 2002, PASJ, 54, 373.
\bibitem[\protect \astroncite{Wang \etal}{2001}]{Nschurch-C2:Wan01}
Wang J.X., Zhou Y.Y., Wang T.G., 1999, ApJ, 523, L129
\bibitem[\protect \astroncite{Wang \etal}{1999}]{Nschurch-C2:Wan99}
Wang J.X., Wang T.G., Zhou Y.Y., 2001, ApJ, 549, 891
\bibitem[\protect \astroncite{Weaver \etal}{1994a}]{Nschurch-C2:Wea94a}
Weaver K.~A., Mushotzky R.~F., Arnaud K.~A., Serlemitsos P.~J., Marshall F.~E., Petre R., Jahoda K.~M., Smale A.~P., Netzer H., 1994a, ApJ, 423, 621
\bibitem[\protect \astroncite{Weaver \etal}{1994b}]{Nschurch-C2:Wea94b}
Weaver K.~A., Yaqoob T., Holt S.~S., Mushotzky R.~F., Matsuoka M., Yamauchi M., 1994b, APJL, 436, L27
\bibitem[\protect \astroncite{Yaqoob \& Warwick}{1991}]{Nschurch-C2:yaq91}
Yaqoob T., Warwick R.~S., 1991, MNRAS, 248, 773
\bibitem[\protect \astroncite{Yaqoob \etal}{1993}]{Nschurch-C2:yaq93}
Yaqoob T., Warwick R.~S., Makino F., Otani C., Sokoloski J.~L., Bond I.~A., Yamauchi M., 1993, MNRAS, 262, 435
\bibitem[\protect \astroncite{Yaqoob \etal}{2001}]{Nschurch-C2:yaq01}
Yaqoob T., Padmanabhan U., Dotani T., George I.~M., Turner T.~J., Weaver K., Nandra K., 2001, Proceedings of the ``X-ray Emission from Accretion onto Black Holes'' Workshop.
\bibitem[\protect \astroncite{Yaqoob \etal}{1995}]{Nschurch-C2:yaq95}
Yaqoob T., Edelson R., Weaver K.~A., Warwick R.~S., Mushotzky R.~F., Serlemitsos P.~J., Holt S.~S, 1995, ApJL, 453, L81
\bibitem[\protect \astroncite{Zdziarski \etal}{1994}]{Nschurch-C2:zdz94}
Zdziarski A.~A., Fabian A.~C., Nandra K., Celotti A., Rees M.~J., Done C., Coppi P.~S., Madejski G.~M., 1994, MNRAS, 269, L55
\bibitem[\protect \astroncite{Zdziarski \etal}{1996}]{Nschurch-C2:zdz96}
Zdziarski A. A., Johnson W. N., Magdziarz P. 1996, MNRAS, 283, 193
\bibitem[\protect \astroncite{Zdziarski \etal}{2000}]{Nschurch-C2:zdz00}
Zdziarski A.~A., Poutanen J., Johnson W.~N., 2000, ApJ, 542, 703
\end{thebibliography}
\end{document}